\documentclass[12pt]{article}
\begin{document}

\title{\bf \Large Momentum and Angular Momentum in the Expanding Universe}

\author{M. Sharif \thanks{e-mail: hasharif@yahoo.com}
\\ Department of Mathematics, University of the Punjab,\\ Quaid-e-Azam
Campus Lahore-54590, PAKISTAN.}

\date{}

\maketitle

\begin{abstract}
A new approach has been used to evaluate the momentum and angular
momentum of the isotropic and homogeneous cosmological models. It
is shown that the results obtained for momentum exactly coincide
with those already available in the literature. However, the
angular momentum expression coincides only for the closed
Friedmann model.
\end{abstract}

\section{Introduction}

There have been various attempts [1-4] to evaluate energy, momentum and
angular momentum in the expanding universe. Rosen [1] and Cooperstock [2]
used the Einstein pseudotensor with Cartesian coordinates to calculate the
total energy of a closed Friedmann Robertson Walker (FRW) universe.

Janusz Garecki [3] used the Pirani [5-7] and Komar solutions [6-8, 9
(chap.11)] to evaluate the energy and other quantities for the isotropic
and homogeneous cosmological models which exist in General Relativity (GR).
The line we wish to follow here uses the Newtonian force concept adapted
to Relativity [10,11]. The idea of re-introducing the Newtonian
gravitational force into the theory of GR [10] arose in an attempt to deal
with the following problem: Gravitation, being non-linear, should dominate
over the Coulomb interaction at some, sufficiently small, scale. At what
scale would it occur? Whereas this question is perfectly valid in
pre-relativistic terms it becomes meaningless in GR. The reason is that
gravitation is expressed in purely geometric terms while electromagnetism
is not. Thus, in Relativity, gravitation possesses a very different status
than the other forces of Nature. Our physical intuition for the other
interactions, nevertheless, rests on the concept of forces. To deal with
gravity and other forces together, we must either express the other forces
geometrically, as in the Kaluza-Klein theories [12], or express gravitation
in the same terms as the other forces. We will follow the latter
alternative as the simpler program to implement.

The plan of the paper is as follows. In the next section we shall briefly
review the essential points of the extended pseudo Newtonian
$(e\psi N)$-formalism for the purpose of application. In section three we
apply the formailsm to the G$\ddot{o}$del universe and evaluate its force,
momentum and angular momentum. In the next section we shall use this
formalism to calculate force and momentum for isotropic and homogeneous
cosmological models. In section five we shall evaluate the angular
momentum for the expanding models.  Finally, in the last section we
summarise the results.

\section{The $e\psi N$-Formalism}

The basis of the formalism is the observation that the tidal force, which
is operationally determinable, can be related to the curvature tensor by

\begin{equation}
F_T^\mu=mR_{\nu\rho\pi}^\mu t^\nu l^\rho t^\pi,\quad (\mu,\nu,\rho,\pi=0,1,2,3),
\end{equation}
where $m$ is the mass of a test particle, $t^\mu=f(x)\delta_0^\mu,
\quad f(x)=(g_{00})^{-1/2}$ and $l^\mu$ is the separation vector. $l^\mu$
can be determined by the requirement that the tidal force have maximum
magnitude in the direction of the separation vector. Choosing a gauge in
which $g_{0i}=0$ (similar to the synchronous coordinate system [13])
in a coordinate basis. We further use Riemann normal coordinates (RNCs) for
the spatial direction, but not for the temporal direction. The reason for
this difference is that both ends of the accelerometer are spatially free,
i.e. both move and do not stay attached to any spatial point. However,
there is a ``memory" of the initial time built into the accelerometer in
that the zero position is fixed then. Any change is registered that way.
Thus ``time" behaves very differently from ``space".

The relativistic analogue of the Newtonian gravitational force called the
$\psi N$ gravitational force, is defined as the quantity whose directional
derivative along the accelerometer, placed along the principal direction,
gives the extremised tidal force and which is zero in the Minkowski space.
Thus the $e\psi N$ force, $F_\mu$, satisfies the equation

\begin{equation}
F_T^{*\mu}=l^\nu F_{;\nu}^\mu ,
\end{equation}
where $F_T^{*\mu}$ is the extremal tidal force corresponding to the
maximum magnitude reading on the dial. Notice that $F_T^{*0}=0$ does not
imply that $F^0=0$. With the appropriate gauge choice and using RNCs
spatially, Eq.(2) can be written in a space and time break up as

\begin{equation}
l^i(F_{,i}^0+\Gamma_{ij}^0F^j)=0,
\end{equation}
\begin{equation}
l^j(F_{,j}^i+\Gamma_{0j}^iF^0)=F_T^{*i}
\end{equation}

A simultaneous solution of the above equations can be found by taking the
ansatz [14]

\begin{equation}
F_0=-m\left[\{\ln (Af)\}_{,0}+g^{ik}g_{jk,0}g^{jl}g_{il,0}/4A\right],
\quad F_i=m(\ln f)_{,i}
\end{equation}
where $A=(\ln \sqrt{-g})_{_{,0}},\quad g=det(g_{_{ij}})$. This force
formula depends on the choice of frame, which is not uniquely fixed.

The new feature of the $e\psi N$ force is its zero component. In special
relativistic terms, which are relevant for discussing forces in a Minkowski
space, the zero component of the four-vector force corresponds to a proper
rate of change of energy of the test particle. Further, we know that in
general an accelerated particle either radiates or absorbs energy according
as $\frac{dE}{dt}$ is less or greater than zero. Thus $F_0$, here, should
also correspond to energy absorption or emission by the background
spacetime. Infact we could have separately anticipated that there should
be energy non-conservation as there is no timelike isometry. In that sense
$F_0$ gives a measure of the extent to which the spacetime lacks isometry.

Another way of interpreting $F_0$ is that it gives measure of the change of
the "gravitational potential energy" in the spacetime. In classical terms,
neglecting this component of the $e\psi N$ force would lead to erroneous
conclusions regarding the "energy content" of the gravitational field.
Cotrariwise, including it enables us to revert to classical cocepts while
dealing with a general relativistically valid treatment. It can be hoped
that this way of looking at energy in relativity might provide a pointer
to the solution of the problem of definition of mass and enrgy in GR.

The spatial component of the $e\psi N$ force $F_i$ is the generalisation of
the force which gives the usual Newtonian force for the Schwarzscild metric
and a $"\frac{Q^2}{r^3}"$ correction to it in the Riesner-Nordstrom metric
[10]. The $\psi N$ force may be regarded as the "Newtonian fiction" which
"explains" the same motion (geodesic) as the "Einsteinian reality" of the
curved spacetime does. We can, thus, translate back to Newtonian terms and
concepts where our intuition may be able to lead us to ask, and answer,
questions that may not have occurred to us in relativistic terms. Notice
that $F_i0$ does not mean deviation from geodesic motion.

The quantity whose proper time derivative is for the test particle. Thus
the momentum four-vector, $p_{_\mu }$, is

\begin{equation}
p_{_\mu }=\int F_\mu dt.
\end{equation}
The spatial components of this vector give the momentum imparted to test
particles as defined in the preferred frame (in which $g_{_{0i}}=0)$.

Consider a ``test rod" (the 1-dimensional extension of a test particle) of
length $\lambda$ in the preferred reference frame. The spin vector is given
[15] by

\begin{equation}
S^\mu=\frac{1}{2}e^{\mu jk\nu}e_{jkl}l^lp_\nu,
\end{equation}
where $e^{\mu\nu\rho\pi}$ is the totally skew fourth rank tensor. In the
preferred frame the spin vector will be proportional to $\it{l}^i$, so that

\begin{equation}
S^i=p_0\it{l}^i.
\end{equation}
(Here the spin can be taken to be negative if the sign of the preferred
direction is reversed.) Taking the magnitude of the spin vector the angular
momentum [16] imparted to the test rod is

\begin{equation}
s=p_0\lambda ,
\end{equation}
where $p_0$ is the zero component of the four-vector momentum. Hence the
physical significance of the zero component of the momentum four-vector
would be that it provides an expression for the spin imparted to a test rod
in an arbitrary spacetime.

\section{Application of the Formalim to the G$\ddot o$del Universe Model}

In 1949, Kurt G$\ddot o$del gave an exact solution of Einstein's field
equations in which the matter takes the form of a pressure-free perfect
fluid ($T_{ab}=\rho u_au_b,$ where $\rho$ is the matter density and $u_a$
the normalised four-vector velocity). The metric can be given [17] in the
form

\begin{equation}
ds^2=-dt^2+dr^2-\frac{1}{2}e^{2ar}d\theta^2+dz^2-2e^{ar}dtd\theta,
\end{equation}
where $a$ is a constant.
The $e\psi N$-force, for the G$\ddot o$del universe model, will become

\begin{equation}
F_0=0,\quad F_1=ma\frac{e^{2ar}}{1-e^{2ar}},\quad F_2=0=F_3
\end{equation}
The corresponding quantities $p_0$ and $p_i$, will turn out to be

\begin{equation}
p_0=constant,\quad p_1=ma\frac{e^{2ar}}{1-e^{2ar}}t+constant,
\quad p_2=constant=p_3
\end{equation}
Thus the spin angular momentum becomes

\begin{equation}
s=constant
\end{equation}

\section{Isotropic and Homogeneous Cosmological Models in GR}

The isotropic and homogeneous cosmological models which exist in the
framework of the GR give us good standard mathematical models of the real
universe, called the Friedmann models. The Friedmann cosmological models
are given by

\begin{equation}
ds^2=dt^2-a^2\left( t\right) \left[ d\chi ^2+\sigma ^2\left( \chi \right)
d\Omega ^2\right] ,
\end{equation}
where $\chi $ is the hyperspherical angle, $\sigma \left( \chi \right) $
is $\sinh \chi ,$ $\chi $ or $\sin \chi $ according as the model is open
$\left(k=-1\right) ,$ flat $\left( k=0\right) $, closed $\left( k=1\right) $
and $ a\left( t\right) $ is the corresponding scale factor. For
matter-dominated Friedmann models, $a(t)$ is given by

\begin{equation}
a(t)=a_0^{1/3}t^{2/3},\quad (k=0),
\end{equation}
\begin{equation}
a(t)=\frac{a_0}{2}(\cosh\eta-1),\quad t=\frac{a_0}{2}(\sinh\eta-\eta),
\quad \eta\ge0,\quad (k=-1),
\end{equation}
\begin{equation}
a(t)=\frac{a_0}{2}(1-\cos\eta),\quad t=\frac{a_0}{2}(\eta-\sin\eta),
\quad 0\le\eta\le2\pi,\quad (k=1),
\end{equation}
where $a_0$ is a constant. The $e\psi N$ force, for the Friedmann models,
is simply

\begin{equation}
F_{_0}=-m\ddot a/\stackrel{.}{a},\qquad F_{_i}=0,
\end{equation}
where a dot denotes differentiation with respect to t. The corresponding
$p_{_0}$ and momentum, $p_{_i},$ imparted to a test particle is

\begin{equation}
p_{_0}=-m\ln \stackrel{.}{a},\qquad p_{_i}=constant.
\end{equation}

For a flat Friedmann model, Eq.$\left( 18\right) $ yields

\begin{equation}
F_{_0}=-m/3t,\qquad F_{_i}=0.
\end{equation}
Thus $F_0$ is proportional to $t^{-1}$ and hence goes to $\infty$ as $t$
approaches to $0$ and it tends to $0$ when $t$ approaches to $\infty$.
Since $F_0$ is negative, it corresponds to the energy dissipation [14]
by the background spacetime.

Consequently, $p_{_0}$ and $p_{_i}$ become

\begin{equation}
p_{_0}=\frac 13m\ln \left( T/t\right) ,\qquad p_{_i}=constant,
\end{equation}
where $T$ is a constant.

For the open Friedmann model, at the arbitrary times, the components of the
$ e\psi N$ force turn out to be

\begin{equation}
F_{_0}=\frac{2m}{a_{_0}\sinh \eta \left( \cosh \eta -1\right) },\qquad
F_{_i}=0.
\end{equation}
Hence $F_0$ goes as $t^{-1}$ for large $t$ as in the case for flat
Friedmann model. The corresponding $p_{_0}$ and $p_{_i}$ are

\begin{equation}
p_{_0}=m\ln \left( \frac{\cosh \eta -1}{\sinh \eta }\right) ,\qquad
p_{_i}=constant.
\end{equation}

For the closed Friedmann universe, the $e\psi N$ force becomes

\begin{equation}
F_{_0}=\frac{2m}{a_{_0}\sin \eta \left( 1-\cos \eta \right) },\qquad
F_{_i}=0.
\end{equation}
This creates a problem at the phase of maximum expansion. However, it can
be resolved by redefining [14] the zero setting of the accelerometer.
This can be explained as follows. The Christoffel symbol $\Gamma_{0j}^i$
appearing in Eq.(4) is zero, for this case, when $F_T^*$ reaches a minimum
value. According to the ansatz used this gives an infinite $e\psi N$ force
at that instant. This is clearly absurd. It was verified that obtaining the
general solution to Eqs.(3) and (4) does not resolve this problem. However,
there is an arbitrariness in what we choose to call the ``zero" of the
accelerometer. There is no a $\it{priori}$ reason to set it at any
particular value. We can then choose to set it at zero at the phase of
maximum expansion, $\eta=\pi$, so as to avoid the infinity in the $e\psi N$
force. Using this resetting, the $e\psi N$ force takes the form

\begin{equation}
F_{_0}=m\frac{4+3\sin ^2\eta +3\cos \eta +\cos ^3\eta }{4a_{_0}\sin \eta
\left( 1-\cos \eta \right) },\qquad F_{_i}=0.
\end{equation}
This gives $F_{_\mu }=0$ at the phase of maximum expansion, $\eta =\pi ,$
of the universe.

The corresponding quantities $p_{_0}$ and $p_{_i}$ are
\begin{equation}
p_0=m\left[\frac{1}{2}\ln \left|\frac{1}{1-\cos \eta }\right|+\frac 38\cos
\eta+\frac 1{16}\sin ^2\eta +c_{_1}\right] ,\qquad p_{_i}=c_{_2}
\end{equation}
where $c_{_1}$and $c_{_2}$are arbitrary constants. Hence, at the phase of
maximum expansion, we have

\begin{equation}
p_{_0}=m\left[ \ln \left( \frac 1{\sqrt{2}}\right) -\frac 38+c_{_1}\right]
,\qquad p_{_i}=c_{_2}
\end{equation}
We see that the momentum, $p_{_i}$, is constant in all the three cases of
the Friedmann universe. Since $c_{_1} $and $c_{_2}$ are arbitrary constants
we can choose them as zero. This choice makes the momentum four-vector,
$ p_{_\mu }$, zero at the phase of maximum expansion.

\section{Angular Momentum in Isotropic and Homogeneous Cosmological Models}

The angular momentum, for the Friedmann models, is

\begin{equation}
s=p_{_0}\lambda =-m\lambda \ln \stackrel{.}{a}
\end{equation}

For a flat Friedmann model, Eq.$\left( 28\right) $ gives

\begin{equation}
s=\frac 13m\lambda \ln \left( T/t\right).
\end{equation}

For the open Friedmann model, at the arbitrary times, the angular momentum
will become

\begin{equation}
s=m\lambda \ln \left( \frac{\cosh \eta -1}{\sinh \eta }\right) .
\end{equation}

For the closed Friedmann universe, the angular momentum takes the form
\begin{equation}
s=m\lambda \left[ \frac 12\ln \left|\frac {1}{1-\cos \eta }\right|+
\frac 38\cos \eta+\frac 1{16}\sin ^2\eta +c_{_1}\right] .
\end{equation}
Hence, at the phase of maximum expansion, we have

\begin{equation}
s=m\lambda \left[ \ln \left( \frac 1{\sqrt{2}}\right) -\frac
38+c_{_1}\right] .
\end{equation}
Since $c_{_1}$ is arbitrary constant. Thus for a suitable choice of this
constant we have the zero spin angular momentum at the phase of maximum
expansion.

\section{Conclusion}

We have applied the $e\psi N$-formalism to the G$\ddot o$del universe
model. Since this model is stationary (non-expanding), the temporal
component of the force must have to be zero which is the requirement of the
procedure. Consequently, the spin angular momentum becomes constant. Also,
the model is only radial coordinate dependent, thus we have the force and
momentum only in the radial direction. As a result, the momentum in the
other directions is constant. It is worth noticing that the results turn
out as the natural requirement of the formalism for non-expanding universe.

Next, we have calculated momentum of a test particle by using $e\psi N
$-formalism. It has been shown that this quantity turns out to be constant
in the free fall rest-frame of the cosmic fluid for the isotropic and
homogeneous relativistic cosmological models. Further we note that the
quantity $p_{_0}$ vanishes for $c_{_1}=\frac 38-\ln \frac 1{\sqrt{2}}$
in a closed Friedmann universe at the phase of maximum expansion and the
quantity $p_{_i}$ can be made zero for $c_{_2}=0$. Thus the momentum
four-vector becomes zero for this particular choice of constants at the
phase of maximum expansion. We have also evaluated spin angular momentum
for the Friedmann universe. This quantity becomes constant for the closed
model. If we choose a particular value of the arbitrary constant $c_{_1}$
as above, we can have the zero spin angular momentum for the closed
universe at the phase of maximum expansion. This is what one would expect
from the formalism. These results coincide with the momentum and angular
momentum results given by Garecki [3]. However, the angular momentum do not
become zero for the flat and open Friedmann models. This seems to be a
problem. To understand the angular momentum imparted to a test rod, a
complete analysis is under progress [18] with reference to gravitational
waves.

\begin{description}
\item  {\bf Acknowledgment}
\end{description}

The author would like to thank Punjab University for the partial financial
support. I am also grateful to the unknown referee for his useful comments.

\end{document}